# Revising Bloom's Taxonomy for Dual-Mode Cognition in Human-AI Systems: The Augmented Cognition Framework

Kayode P. Ayodele[1], Enoruwa Obayiuwana[1], Aderonke R. Lawal[2], Ayorinde Bamimore[3], Funmilayo B. Offiong[4], Emmanuel A. Peter[1]

[1] Department of Electronic and Electrical Engineering, Obafemi Awolowo University, Nigeria
[2] Department of Computer Science and Engineering, Obafemi Awolowo University, Nigeria
[3] Department of Chemical Engineering, Obafemi Awolowo University, Nigeria
[4] Department of Engineering, Glasgow Caledonian University, Scotland

## Abstract

As artificial intelligence (AI) models become routinely integrated into knowledge work, cognitive acts increasingly occur in two distinct modes: individually, using biological resources alone, or distributed across a human–AI system. Existing revisions to Bloom's Taxonomy treat AI as an external capability to be mapped against human cognition rather than as a driver of this dual-mode structure, and thus fail to specify distinct learning outcomes and assessment targets for each mode. This paper proposes the Augmented Cognition Framework (ACF), a restructured taxonomy built on three principles. First, each traditional Bloom level operates in two modes (Individual and Distributed) with mode-specific cognitive verbs. Second, an asymmetric dependency relationship holds wherein effective Distributed cognition typically requires Individual cognitive foundations, though structured scaffolding can in some cases reverse this sequence. Third, a seventh level, Orchestration, specifies a governance capacity for managing mode-switching, trust calibration, and partnership optimization. We systematically compare existing AI-revised taxonomies against explicit assessment-utility criteria and show, across the frameworks reviewed, that ACF uniquely generates assessable learning outcomes for individual cognition, distributed cognition, and mode-governance as distinct targets. The framework addresses fluent incompetence, the central pedagogical risk of the AI era, by making the dependency relationship structurally explicit while accommodating legitimate scaffolding approaches.

# 1. Introduction

With increasing ubiquity of artificial intelligence (AI) models, cognition has undergone a structural transformation. Every cognitive act, from basic recall to complex creation, can now be performed in two fundamentally different modes: individually, using biological cognitive resources, or distributed across a human-AI system. Consider a physician diagnosing a complex case. She can reason through the differential diagnosis using her training, experience, and biological memory. Or, she can engage an AI system to surface patterns across millions of cases, propose diagnostic hypotheses, and evaluate treatment options: distributed cognition. These are not the same cognitive act performed with and without assistance. They are qualitatively different cognitive processes requiring different competencies, producing different failure modes, and demanding different assessments. Every domain of intellectual work now increasingly admits such dual-mode cognition. The professional who cannot operate effectively in both modes, and who cannot judge when each is appropriate, will be cognitively incomplete by contemporary standards.

Bloom's Taxonomy (Bloom et al, 1956), the dominant framework for specifying educational objectives since 1956, was designed for a world of single-mode cognition. Its six levels describe cognitive processes performed by individual minds using biological resources. The emergence of AI capable of participating in cognition across all six levels does not merely add new skills to be learned; it reveals that cognition itself now has a dimension the taxonomy does not capture.

This paper argues that educational taxonomy must reflect the emerging multidimensional structure of contemporary cognition. We propose the Augmented Cognition Framework (ACF), in which each traditional level operates in two modes (Individual and Distributed). We also propose a seventh level, Orchestration, governing mode selection and partnership management.

The manuscript proceeds as follows: Section 2 sets out the conceptual foundation for distributed cognition; Section 3 motivates its educational salience; Section 4 defines the assessment-utility lens for reviewing existing taxonomies; Sections 5–7 present the ACF, its threshold logic, and implications for curriculum and assessment; Section 8 reports a comparative evaluation; Section 9 outlines limitations and future research.

## 2. AI as Cognitive Partner: Philosophical Foundation

The dominant current metaphor for AI in education is the tool. Like calculators, word processors, and search engines, AI systems are conceptualized as instruments that extend human capabilities while leaving the structure of cognition intact. With the rise and ubiquitous access to capable generative AI models, we believe this metaphor has now become inadequate. It obscures a categorical difference that determines how AI should be treated pedagogically.

### 2.1 Extended Mind Basis for Post-Tool Status of AI

Previous technologies amplify specific human capacities while preserving the architecture of cognition. A calculator performs arithmetic, but the human decides what to calculate and interprets results. AI is categorically different. When a professional asks a large language model (LLM) to analyse a dataset or propose solutions to a design problem, the AI does not merely amplify a specific capacity. It participates in reasoning itself. It generates inferences, constructs arguments, identifies patterns, weighs considerations, and proposes conclusions. The human interacts with these outputs not as raw data to be processed but as cognitive contributions to be evaluated, refined, or rejected. The distinction can be stated formally: tools operate on the objects of cognition; cognitive partners participate in cognition as process.

Drawing on Clark and Chalmers' (1998) extended mind thesis, we propose that AI qualifies as a cognitive partner when human-AI interaction meets four coupling conditions:

i. Reliability: The AI is consistently available and accessible when needed for cognitive work. The professional routinely turns to the AI as part of their cognitive workflow.

ii. Integration: AI outputs are directly incorporated into reasoning and action without requiring complete independent verification. The human treats AI contributions as inputs to further cognitive processing, not as external data requiring full reconstruction.

iii. Automatic endorsement: Within appropriate domains and confidence levels, AI outputs are accepted as prima facie credible, subject to override but not requiring active validation for every claim. This parallels how one treats one's own memory: generally trusted, but occasionally checked.

iv. Reciprocal adaptation: The human adapts their cognitive processes to leverage AI capabilities, while AI outputs are shaped by human direction. The interaction is bidirectional, with each component influencing the other's contributions.

When these conditions are met, the human-AI system functions as a cognitive unit. The question 'did the human or the AI think that?' becomes as philosophically arbitrary as 'did the hippocampus or the prefrontal cortex think that?' The relevant unit of analysis becomes the coupled system, not its components.

Critics may object that existing software such as statistical packages, theorem provers, expert systems also 'participate' in reasoning by these criteria. This objection has merit only as a matter of degree: the tool-partner distinction is better understood as a continuum than a binary. However, generative AI occupies a qualitatively different position on this continuum. Unlike statistical packages that execute specified algorithms, LLMs generate novel inferences, construct arguments not explicitly programmed, and produce outputs that require evaluative engagement rather than mere acceptance or rejection of a calculation. The coupling criteria are met more fully, and the pedagogical implications are correspondingly more profound.

## 2.2 AI as an Exocortex

When used in cognition-augmenting mode, we conceptualize AI as an 'exocortex': an external cognitive resource that, when properly coupled with biological cognition, extends what the human-AI system can perceive, remember, and think. The term acknowledges that AI functions as a cognitive substrate without implying neural homology or anthropomorphic properties. AI is not a brain; it is a functionally integrated cognitive resource that operates alongside biological cognition.

A caution is warranted. The exocortex framing, while more precise than 'tool,' carries its own risks. Students may overestimate AI reliability or attribute to it forms of understanding it does not possess. This risk is real but is directly addressed by the framework's Orchestration level, which makes trust calibration and degradation detection explicit competencies. The framing enables clear thinking about human-AI cognition; Orchestration ensures students learn to manage the partnership appropriately.

The exocortex framing has three advantages over the tool metaphor. First, it acknowledges that AI participates in cognition rather than merely supporting it. The quality of human-AI coupling matters: just as different brains reason with different effectiveness, different patterns of human-AI interaction produce different cognitive outcomes.

Second, it implies that learning to work with AI is not merely acquiring a skill (like learning to use a calculator) but developing a form of cognition (like learning a language or mathematical notation). Writing transformed not just what humans could record but how they could think—externalized language enabled new forms of abstract reasoning. AI represents a comparable cognitive transformation.

Third, it suggests appropriate pedagogical responses. If AI is a tool, educators should teach tool operation and maintenance of tool-independent competencies. If AI is a cognitive partner, educators must develop students' capacities for effective distributed cognition; cognition that occurs across the human-AI system rather than within the human alone.

## 3. The Empirical Inevitability of Distributed Cognition

The foregoing philosophical argument established that AI can function as cognitive partner. An empirical argument will now be made to establish that it will: distributed cognition will become the dominant mode of professional intellectual work, making its inclusion in educational taxonomies practically necessary regardless of one's philosophical commitments.

The World Economic Forum's Future of Jobs Report 2025 projects that AI will displace 92 million jobs while creating 170 million new ones (World Economic Forum, 2025). These new roles are not simply 'AI jobs'. They are existing roles transformed by AI integration. The International Monetary Fund (Cazzaniga et al, 2024) notes that nearly 40% of global jobs are 'exposed' to AI-driven change, while emphasizing that students now entering higher education need 'cognitive, creative, and technical skills that complement AI and help them use it rather than compete with it.'

The Indeed Hiring Lab's AI at Work Report (2025) provides granular evidence. Using their GenAI Skill Transformation Index, they find that 26% of jobs posted could be 'highly transformed' by GenAI, with another 54% 'moderately transformed.' Almost half (46%) of skills in a typical job posting are poised for 'hybrid transformation,' where human oversight remains critical but AI performs significant routine cognitive work.

Professionals who effectively leverage AI will outperform those who do not. Organizations that integrate AI into knowledge work will have productivity advantages over those relying solely on human cognition. A student entering university in 2026 will graduate around 2030 and work until approximately 2065. Their career will unfold almost entirely in a world where distributed cognition is the professional norm. Educational taxonomy must reflect this reality. Curricula designed around individual cognition alone prepare students for a world that is rapidly ceasing to exist.

## 4. Systematic Comparison of Existing AI-Revised Taxonomies

To compare AI-revised Bloom frameworks, we apply a single meta-criterion: assessment utility, which is the extent to which a framework yields explicit, observable learning outcomes. We judge utility across three assessment targets: (i) Unaided biological cognition, (ii) human–AI cognition expressed through assessable verbs, and (iii) orchestration: Governing the partnership through mode selection, trust calibration, and management/verification of AI contributions.

### 4.1 Framework Descriptions

We identified nine relevant works through structured search, selecting frameworks that explicitly engage with Bloom's Taxonomy in AI-affected learning contexts. These range from practitioner guidance documents to proposed taxonomic revisions to systematic reviews of AI-related educational technologies.

   i. Oregon State University's Bloom's Taxonomy Revisited (Oregon State University Ecampus, 2024): This framework preserves the six traditional levels and creates a dual mapping between "Distinctive Human Skills" and "How GenAI Can Supplement Learning Processes" at each level. It draws on the MAGE framework for assessing AI's critical thinking capabilities (Zaphir et al., 2024) and provides practical guidance for faculty reflection on course activities and assessments.

   ii. Gonsalves (2024): This framework adds AI-specific competencies: "melioration" (selecting and integrating appropriate tools), "ethical reasoning," "interrogating AI outputs," and "articulating precise prompts", as supplements to the existing Bloom structure. It argues that Bloom's taxonomy fails to address the cognitive demands of AI-assisted learning and the solution is a revised framework incorporating AI-specific competencies.

iii. AIEd Bloom's Taxonomy (Hmoud & Shaqour, 2024): This framework replaces traditional Bloom levels with six AI-aligned levels: Collect, Adapt, Simulate, Process, Evaluate, and Innovate. Each level describes what AI platforms can do to support learning, with specific tool recommendations for each level.

iv. LBET (Luo et al., 2025): The LLM-driven Bloom's Educational Taxonomy delineates two stages (Exploration & Action and Creation & Metacognition) subdivided into seven phases: Perceiving, Searching, Reasoning, Interacting, Evaluating, Organising, and Curating. It is specifically designed for information literacy with LLMs.

v. Bloom Meets Gen AI (Jain & Samuel, 2025): This framework proposes viewing cognitive skills as "interconnected nodes rather than hierarchical steps." It introduces new categories including "Ventriloquising" (passive replication of AI-retrieved information) and "Co-curating" (collaborative content generation with AI), and collapses Understanding, Applying, Analysing, and Evaluating into "Critical Understanding." It explicitly adopts "Intelligence Augmentation" as its conceptual foundation.

vi. Nehru et al. (2025): This work applies the Revised Bloom's Taxonomy (Anderson and Krathwohl, 2001) to AI-digital and propose a multi-phase approach for aligning Revised Bloom's levels with AI-enabled digital tools and online learning activities, illustrating how tools can be selected to support objectives across the hierarchy.

vii. AlAfnan (2024): This framework proposes a new six-category taxonomy for the AI era: (1) Knowledge and Comprehension, (2) Synthesis and Evaluation, (3) Ethical and Moral Reasoning, (4) Application and Strategic Thinking, (5) Creativity and Innovation, and (6) Lifelong Learning and Adaptability. It integrates Education for Sustainable Development (ESD) goals alongside AI considerations.

viii. Hammad et al. (2024): Using the PRISMA framework, this systematic review proposes a taxonomy of AI-based assessment educational technologies with seven categories: Proctoring, Auto Grading, Adaptive Assessment, Feedback, Immersive Assessment, Gamification, and Learner Behaviour. It classifies 24 research papers and 68 tools.

ix. Almatrafi & Johri (2025): This work investigates using GPT-4 to automatically classify course learning outcomes according to Bloom's Taxonomy levels. It tests multiple prompt engineering strategies on a dataset of 1,000 annotated learning outcomes, achieving classification accuracy comparable to human raters.

## 4.2 The Dimensionality Gap

Existing Bloom–AI adaptations exhibit a shared structural limitation that warrants taxonomic revision rather than incremental extension. Across board, AI is treated as an external tool to be mapped onto human cognition, instead of as a condition that has introduced two qualitatively different modes of cognition: individual (biological) and distributed (human–AI).

This limitation appears in three recurrent design elements. First, parallel mapping preserves Bloom's hierarchy while aligning AI capabilities with each level (e.g., dual-column layouts; tool-to-level mappings). These frameworks answer what AI can do at each level, but they do not specify distinct, assessable learning outcomes for cognition enacted in distributed systems; "analysing with AI" remains indistinguishable from "analysing without AI." Second, additive supplementation inserts AI-specific skills (e.g., prompt articulation, ethical reasoning, co-curation) into the existing structure. While pedagogically valuable, these skills operate as appendices; the underlying unit of analysis remains the individual learner, and the taxonomy does not redefine how each cognitive level changes under distributed performance. Third, platform-functional taxonomies replace cognitive processes with AI functions or assessment technologies (e.g., Collect/Adapt/Simulate). Such schemes may aid instructional design, but they cannot generate human learning outcomes because they classify technological capabilities rather than cognition.

Even conceptually advanced proposals that acknowledge altered cognitive acts (e.g., terms distinguishing reproduction from collaboration) typically treat these as isolated additions; abandoning hierarchy also weakens the sequencing logic required for curriculum design. The core problem is therefore precise: current frameworks preserve individual cognition as the default and position AI as an ancillary complication: something to benchmark, add skills for, or constrain, rather than as a basis for re-specifying cognition itself.

A multidimensional taxonomy must instead define competencies for effective performance in both individual and distributed modes, and for governing mode selection. This requires reinterpreting each cognitive level across two modes and treating orchestration as a distinct, assessable competence. We therefore propose the Augmented Cognition Framework as a structural revision that makes dual-mode cognition explicit. Section 8 reports a comparative evaluation using design-science taxonomy criteria; the assessment-utility dimensions introduced here inform the Completeness and Appropriateness-to-Purpose judgments.

## 5. The Augmented Cognition Framework

The Augmented Cognition Framework (ACF) as presented in Figure 1, is built on three structural principles designed to address the limitations identified above.

### 5.1 Structural Principles

**Principle 1: Dual-Mode Operation.** Each traditional Bloom level now operates in two modes (Individual and Distributed) with mode-specific cognitive verbs. Individual cognition operates as in traditional Bloom's: the human performs cognitive work using biological resources. Distributed cognition involves the human-AI system: cognitive work occurs across human and artificial agents, with the human's distinctive contribution being direction, evaluation, and integration rather than solo performance. The Distributed mode is not conceived as 'doing X with AI help'; it is a transformed cognitive act with its own characteristic verb.

**Principle 2: Asymmetric Dependency with Scaffolding Provisions.** Effective distributed cognition typically depends on individual cognitive foundations. The relationship is asymmetric in the default case: Individual grounds Distributed. One cannot curate information without enough domain knowledge to recognise errors; one cannot discriminate between genuine and superficial AI explanations without understanding deep enough to detect the difference. However, this dependency admits exceptions under structured scaffolding conditions, detailed in Section 6.2. The key diagnostic is not sequence but outcome: can the student eventually perform without AI assistance when required? If not, fluent incompetence has occurred regardless of which mode was practiced first.

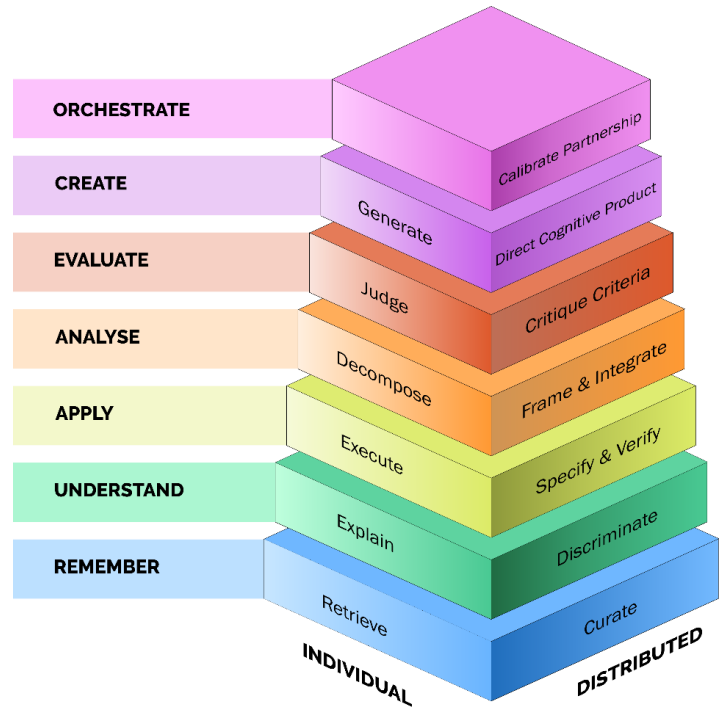

Figure 1: The Augmented Cognition Framework

**Principle 3: Orchestration as Meta-Level.** A seventh level (Orchestration) has been introduced to govern the meta-cognitive capacity to manage human-AI cognitive partnership. Orchestration exists only in distributed cognition; there is no individual-mode equivalent. It is inherently about managing the relationship between modes.

## 5.2 Transforming the Traditional 6 Levels for the Age of AI

The traditional six levels are re-specified across two modes. Distributed performance presupposes the corresponding individual competence; without it, learners may produce fluent outputs that are unreliable or unaccountable.

### 5.2.1. Level 1: Remember

**Individual Mode (Verb: Retrieve):** The traditional cognitive process of accessing stored information from memory. Some information must be immediately available without external consultation, and the process of encoding information builds understanding.

**Distributed Mode (Verb: Curate):** When AI handles retrieval, the human contribution shifts to validation and selection. Can the student recognise hallucinated content? Do they know what information to request? Can they judge relevance and reliability? Curation requires enough domain knowledge to evaluate AI retrieval without necessarily holding all information in biological memory.

### 5.2.2. Level 2: Understand

**Individual Mode (Verb = Explain):** Constructing meaning through interpretation, exemplification, classification, summarization, inference, and comparison. Understanding manifests as the ability to articulate concepts in one's own words and generate novel examples.

**Distributed Mode (Verb = Discriminate):** AI generates fluent explanations that may or may not reflect genuine comprehension. It can produce text that sounds like understanding without any underlying grasp. The human contribution becomes discrimination: distinguishing genuine insight from sophisticated mimicry, in AI outputs and, critically, in oneself. This has two components: (a) Audit: evaluating external AI outputs for genuine versus superficial understanding; (b) Self-diagnose: recognizing when one's own apparent understanding is actually borrowed from AI without genuine comprehension.

### 5.2.3. Level 3: Apply

**Individual Mode (Verb = Execute):** Carrying out procedures, implementing methods, solving problems using learned approaches. Competence manifests through correct execution in novel situations.

**Distributed Mode (Verb = Specify & Verify):** When AI handles execution (writing code, performing calculations, generating documents), humans must specify what procedure to apply and verify that execution was appropriate. This requires understanding procedures well enough to recognise correct and incorrect application without necessarily executing at production speed oneself.

### 5.2.4. Level 4: Analyse

**Individual Mode (Verb = Decompose):** Breaking material into parts, detecting structure, differentiating relevant from irrelevant, organising components, attributing underlying perspectives. Analysis reveals what is not apparent on the surface.

**Distributed Mode (Verb = Frame & Integrate):** AI can decompose with impressive thoroughness, but humans must frame what to analyse and integrate AI analysis with contextual knowledge AI lacks. What question are we actually asking? What context shapes interpretation? How do analytical findings connect to the broader situation? What has the AI missed because it lacks access to unstated context?

### 5.2.5. Level 5: Evaluate

**Individual Mode (Verb = Judge):** Checking against standards, critiquing based on criteria, assessing quality and appropriateness. Judgment involves applying values and standards to reach warranted conclusions about worth.

**Distributed Mode (Verb = Critique Criteria):** AI can judge against stated criteria with remarkable consistency. But humans must critique whether the criteria capture what actually matters. A contract might meet all specified requirements yet fail to serve the client's actual interests. A treatment might satisfy clinical guidelines yet be inappropriate for this patient's values. The distinctively human contribution is meta-evaluation: judging the adequacy of the criteria themselves.

### 5.2.6. Level 6: Create

**Individual Mode (Verb = Generate):** Putting elements together to form novel wholes, reorganising into new patterns, producing original work. Creation involves bringing something new into existence that reflects the creator's intent.

**Distributed Mode (Verb= Direct Cognitive Product):** AI generates content in abundance, including text, images, code, designs. The human contribution in this process becomes directing the cognitive product: specifying the problem the creation solves, defining constraints imposed by audience and context, and establishing observable success criteria. This is operationalized as: (a) Problem specification: articulating the cognitive or practical problem the creation addresses; (b)

Constraint definition: identifying requirements imposed by audience, context, and purpose; (c) Success criteria: stating observable indicators that the creation achieves its aims. This framing keeps the competency within the cognitive domain while acknowledging that creation is inherently purposive.

## 5.3 The Seventh Level: Orchestration

Above the six traditional levels, we propose a seventh: Orchestration. This is the meta-cognitive capacity to govern the human-AI cognitive partnership itself. Orchestration exists only in distributed cognition. it is inherently about managing the relationship between individual and distributed modes.

A critical distinction must be made between Orchestration and traditional metacognition. Classic metacognition, as developed in the self-regulated learning literature, involves monitoring and controlling one's own cognitive processes, including planning, monitoring comprehension, evaluating progress, and adjusting strategies. The target of classic metacognition is internal: one's own cognition.

Orchestration involves a categorically different target: managing cognition distributed across a human-AI system where one component (the AI) is stochastic, partially opaque, and capable of confident errors. This is not self-monitoring but partnership monitoring. The challenges are distinct:

In classic metacognition, you monitor a system (your own mind) that you have privileged access to, that operates according to patterns you have learned over a lifetime, and that fails in ways correlated with your own sense of difficulty. In Orchestration, you monitor a system (the AI) that you have no privileged access to, that operates according to patterns you cannot fully observe, and that fails in ways uncorrelated (or inversely correlated) with apparent difficulty. The AI may be confidently wrong on easy problems and surprisingly right on hard ones. Calibrating trust requires different cognitive work than monitoring comprehension.

Orchestration involves four core capacities:

i. Mode-switching judgment: When should I rely on AI versus work individually? When is AI assistance enhancing cognition versus degrading judgment? What types of tasks benefit from distribution versus require individual engagement?

ii. Trust calibration: How reliable is AI for this type of task in this domain? When should I verify versus accept? How do I maintain appropriate scepticism without inefficient redundancy? Trust must be calibrated to context: AI excels at some tasks and fails at others, and the boundaries shift constantly.

iii. Degradation detection: Am I becoming over-reliant? Is my individual capacity atrophying from disuse? Am I deferring in situations where I should exercise independent judgment? The orchestrator monitors their own cognitive state and recognises when distributed cognition is substituting for rather than augmenting individual cognition.

iv. Partnership optimization: How do I structure my interaction with AI to get the best cognitive outcomes? What prompting strategies, workflows, and verification procedures maximize the quality of distributed cognition?

Orchestration is the distinctive competency of the AI era. It distinguishes effective from ineffective distributed cognition. A professional who cannot orchestrate will either over-rely on AI (losing the individual cognitive foundation that makes their contribution valuable) or under-rely (losing the productivity gains that distributed cognition enables).

**5.4 Summary Framework**

Table 1 summarises the Augmented Cognition Framework by pairing Individual- and Distributed-mode verbs at each level and stating the dependency claim.

# 6. Fluent Incompetence, Thresholds, and Scaffolding

The dependency principle (that Distributed cognition typically requires Individual foundations) addresses the central pedagogical risk of the AI era: fluent incompetence. This is the ability to produce sophisticated outputs without genuine understanding, enabled by AI's capacity to generate polished work regardless of the user's cognitive state.

*Table 1: The Augmented Cognition Framework*

| Level | Individual Mode | Distributed Mode | Dependency |
|---|---|---|---|
| 1. Remember | Retrieve | Curate | Cannot curate without knowledge to recognise errors |
| 2. Understand | Explain | Discriminate (Audit + Self-diagnose) | Cannot discriminate without understanding to detect superficiality |
| 3. Apply | Execute | Specify & verify | Cannot verify without execution experience to recognise misapplication |
| 4. Analyse | Decompose | Frame & integrate | Cannot frame without analytical capacity to specify what matters |
| 5. Evaluate | Judge | Critique Criteria | Cannot critique criteria without judgment to assess proper application |
| 6. Create | Generate | Direct Cognitive Product | Cannot direct effectively without generative experience |
| **7. Orchestrate** | — | Mode-switch, Trust-calibrate, detect degradation, Optimize partnership | Requires foundations across all levels to govern mode selection |

Fluent incompetence is dangerous precisely because it looks like competence to outside observers and often to the students themselves. AI-assisted outputs can be polished, sophisticated, and superficially correct while lacking the understanding that would allow the student to recognise errors, adapt to novel situations, or exercise genuine judgment. The student passes assessments, receives credentials, and enters professional practice without the cognitive foundations that make professional judgment possible.

**6.1 Observable Thresholds for Dependency**

The dependency principle requires operationalization. Stating that 'Individual competence must precede Distributed practice' is philosophically correct but pedagogically vague. What counts as sufficient Individual competence? We propose minimum viable individual competence be specified through observable diagnostics at each level:

**Level 1 (Remember leads to Curate):** The student can identify factual errors seeded into AI-retrieved information at a rate significantly above chance. Diagnostic: Present AI-generated summaries with 3-5 deliberate factual errors; student must identify majority without external reference.

**Level 2 (Understand leads to Discriminate):** The student can generate counterexamples to AI explanations that are superficially plausible but conceptually flawed. Diagnostic: Present AI explanation with subtle conceptual error; student must identify error and explain why it fails.

**Level 3 (Apply leads to Specify & Verify):** The student can predict failure modes of procedures before applying them and identify when AI has applied the wrong procedure to a given problem. Diagnostic: Present AI-executed procedure with category error (right technique, wrong problem type); student must identify misapplication.

**Level 4 (Analyse leads to Frame & Integrate):** The student can identify what contextual factors AI analysis has missed and reframe analysis questions to surface different aspects. Diagnostic: Present AI analysis of case; student must identify unstated contextual factors that would change interpretation.

**Level 5 (Evaluate leads to Critique Criteria):** The student can articulate what evaluation criteria fail to capture and propose modified criteria for edge cases. Diagnostic: Present AI evaluation that satisfies stated criteria but fails on unstated important dimension; student must identify the gap.

**Level 6 (Create leads to Direct Cognitive Product):** The student can distinguish AI-generated content that solves the specified problem from content that is superficially impressive but misses the target, and articulate the difference in terms of problem specification, constraints, and success criteria. Diagnostic: Present two AI-generated artifacts addressing the same brief; student must identify which better satisfies the cognitive task and explain why.

These diagnostics make the dependency principle falsifiable. A curriculum can be tested: do students who pass the Individual threshold diagnostics before practicing Distributed mode show better long-term outcomes than those who do not? The framework becomes empirically tractable rather than merely conceptually appealing.

We note that precise passing thresholds (e.g., 'identify 4 of 5 errors' versus '3 of 5') require empirical calibration that will vary by discipline and context. The framework provides the structure for such calibration; determining optimal thresholds is future empirical work.

## 6.2 The Scaffolding Exception: When Distributed Can Precede Individual

The dependency principle as stated (Individual typically precedes Distributed) admits exceptions. Emerging research on AI-assisted learning suggests that under specific conditions, structured distributed practice can scaffold the development of individual competence rather than merely requiring it as a precondition.

Consider a student struggling with basic comprehension in a domain. AI-generated explanations, worked examples, and interactive feedback may help the student build the internal schemas that constitute individual understanding. In such cases, distributed cognition functions as scaffolding for individual cognition rather than depending on it.

We therefore articulate three conditions under which the typical sequence may be reversed:

i. Deliberate scaffolding design: The distributed practice is explicitly structured to build individual competence, not to substitute for it. The AI interaction is designed with fading: progressive reduction of AI support as student competence develops.

ii. Verified transfer: The student must eventually demonstrate individual mode competence without AI assistance. If the student cannot perform without AI after scaffolded practice, fluent incompetence has occurred regardless of sequence. Transfer to unassisted performance is the ultimate diagnostic.

iii. Metacognitive awareness: The student understands that AI assistance is temporary scaffolding, not permanent support. They are working toward unassisted competence, not settling into permanent dependence.

Under these conditions, the revised dependency principle becomes: Individual competence is required for effective Distributed practice as the default, but structured scaffolding that verifiably builds Individual competence can reverse the sequence. The key diagnostic is outcome, not sequence: can the student eventually perform without AI assistance when required?

This revision accommodates legitimate pedagogical uses of AI scaffolding while preserving the framework's core insight: distributed cognition that does not eventually produce individual competence is not augmentation but substitution, and substitution produces fluent incompetence.

## 7. Implications for Curriculum Design and Assessment

### 7.1 Writing Learning Outcomes

The framework enables educators to write learning outcomes for both modes at each level. Consider examples:

**Level 3 (Apply) in Accounting:**

*Individual outcome:* Students will calculate appropriate depreciation schedules for capital assets using straight-line and declining-balance methods.

*Distributed outcome:* Students will verify AI-generated depreciation schedules, identify computational errors, and assess whether the chosen method is appropriate given asset type and applicable standards.

**Level 4 (Analyse) in Law:**

*Individual outcome:* Students will identify key legal issues in a contract dispute and structure analysis of each.

*Distributed outcome:* Students will frame contract disputes for AI analysis by specifying relevant jurisdictions, precedents, and contextual factors, then integrate AI analysis with client-specific circumstances the AI cannot access.

Both outcome types matter. The Individual outcome builds cognitive foundation; the Distributed outcome prepares for professional practice. Curricula should include both, with appropriate sequencing respecting the dependency relationship (or its scaffolding exception).

### 7.2 Assessment Design: A Worked Example

To demonstrate the framework's assessment utility, we present a worked example with paired rubrics for a single task context.

**Task Context:** Analysing a company's financial health for investment recommendation (Level 4: Analyse)

**7.2.1. Individual Mode Assessment (Decompose):**

Students receive financial statements without AI access and must: identify key financial ratios relevant to investment decisions; decompose performance into component factors (revenue growth, margin trends, capital efficiency); detect anomalies or inconsistencies requiring investigation; structure findings into coherent analytical narrative.

*Rubric dimensions:* Comprehensiveness of ratio identification; accuracy of calculations; quality of decomposition logic; identification of non-obvious patterns.

**7.2.2. Distributed Mode Assessment (Frame & Integrate):**

Students receive same financial statements plus AI analysis tools and must: frame analytical questions for AI (what to analyse, what comparisons to make); evaluate AI-generated analysis for completeness and accuracy; identify contextual factors AI analysis missed (industry-specific considerations, recent management changes, pending litigation mentioned in footnotes); integrate AI analysis with contextual knowledge into recommendation.

*Rubric dimensions:* Quality of analytical framing; identification of AI analysis gaps; integration of context AI cannot access; justified departure from or acceptance of AI conclusions.

**7.2.3. Orchestration Assessment:**

Students submit reflection addressing: Which aspects of this analysis benefited most from AI assistance and why? Where did you override AI conclusions and on what basis? What would you do differently if AI were unavailable? How did you verify AI outputs and was verification sufficient? What aspects of your individual analytical capacity did this task exercise versus potentially atrophy?

*Rubric dimensions:* Accuracy of mode-switching rationale; appropriateness of trust calibration; evidence of degradation awareness; sophistication of partnership optimization reasoning.

This tri-rubric structure (Individual, Distributed, Orchestration) is what the ACF uniquely generates. No other surveyed framework provides the architecture for this assessment design.

### 7.3 Implementation Flexibility

A legitimate concern about the tri-rubric structure is implementation burden. Developing and applying three distinct rubrics for every assessment could significantly increase faculty workload, particularly in high-enrolment or under-resourced contexts.

We therefore clarify: the tri-rubric structure represents the framework's full assessment capability, not a requirement for every assignment. Programs may implement the framework flexibly:

**Mode separation across assignments:** Individual and Distributed modes may be assessed in separate assignments rather than combined. A midterm might assess Individual competence; a project might assess Distributed competence; a final reflection might assess Orchestration.

**Sampling strategies:** Not every assignment requires all three rubrics. A course might fully assess all three modes only for capstone work, with earlier assignments focusing on one mode at a time.

**Progressive introduction:** Early courses may focus primarily on Individual mode with limited Distributed exposure. Advanced courses shift toward Distributed and Orchestration assessment. The full framework applies across a program, not necessarily within each course.

**Orchestration as periodic audit:** Orchestration assessment need not accompany every Distributed mode task. A semester might include 2-3 dedicated Orchestration reflections rather than requiring reflection on every AI-assisted assignment.

The framework enables comprehensive assessment design; educators retain discretion over implementation intensity appropriate to their context.

### 7.4 Curriculum Sequencing

The dependency relationship suggests clear sequencing principles. At each cognitive level, students should typically develop Individual mode competency (verified by threshold diagnostics) before extensive Distributed mode practice. The scaffolding exception applies when distributed practice is explicitly designed to build individual competence with verified transfer.

For introductory courses, emphasis should fall on Individual mode competencies with limited, structured AI integration (potentially as scaffolding). For advanced courses, emphasis can shift toward Distributed mode competencies and Orchestration, but with continued assessment of Individual foundations. For professional preparation, Orchestration becomes central, with explicit attention to mode-switching judgment in authentic professional contexts.

## 8. Systematic Comparative Evaluation

Having presented the ACF and its implications for curriculum design and assessment, we now provide a systematic comparison against existing frameworks using established taxonomy evaluation criteria. This evaluation substantiates the dimensionality gap claim introduced in Section 4.2 and demonstrates ACF's unique contribution.

### 8.1 Evaluation Methodology

Design-science work on taxonomy development emphasises that evaluation criteria should be explicit and derived from a stated meta-characteristic (Nickerson et al., 2013). In this study, we operationalise assessment-utility evaluation by scoring each framework on seven criteria:

i. Lucidity (LU): Clarity and interpretability of categories for consistent application

ii. Orthogonality (OR): Non-overlap and mutual exclusivity of categories

iii. Completeness (CO): Coverage of the domain specified by the meta-characteristic

iv. Parsimony (PA): Parsimony—absence of redundant categories

v. Appropriateness-to-Purpose (AP): Suitability for generating assessable learning outcomes

vi. Generality (GE): Transferability across disciplines and contexts

vii. Evidence of Usefulness (EV): Documented validation or adoption

Our meta-characteristic is: taxonomy of learning outcomes for specifying and assessing cognitive objectives in AI-affected learning environments. This privileges frameworks that generate assessable learning outcomes over those that classify technologies or provide instructional guidance without outcome specification.

Each criterion was scored on a 0–4 scale (0 = absent/fails; 1 = minimal; 2 = partial; 3 = substantial; 4 = comprehensive). Supplementary Materials A provides the full scoring rubric (criterion-specific anchors), the extracted evidentiary excerpts for each framework, and the resulting score justifications, enabling audit and replication.

## 8.2 Comparative Results

Table 2 presents evaluation scores. Original Bloom's Taxonomy (Anderson & Krathwohl, 2001) is included as baseline.

Table 2: Systematic Evaluation Against Design-Science Taxonomy Criteria

| Framework | LU | OR | CO | PA | AP | GE | EV | Total |
|---|---|---|---|---|---|---|---|---|
| Original Bloom (2001) | 4 | 3 | 2 | 4 | 3 | 4 | 4 | **24** |
| ACF (Proposed) | 4 | 3 | 4 | 3 | 4 | 3 | 1 | **22** |
| Oregon State (2024) | 4 | 3 | 2 | 4 | 2 | 4 | 2 | **21** |
| Gonsalves (2024) | 3 | 2 | 3 | 3 | 3 | 3 | 2 | **19** |
| Nehru et al. (2025) | 3 | 3 | 2 | 4 | 2 | 3 | 1 | **18** |
| AlAfnan (2024) | 3 | 2 | 2 | 3 | 2 | 2 | 1 | **15** |
| Jain & Samuel (2025) | 2 | 1 | 3 | 2 | 2 | 3 | 1 | **14** |
| LBET (Luo et al., 2025) | 3 | 2 | 2 | 2 | 2 | 1 | 1 | **13** |
| Hammad et al. (2024) | 3 | 3 | 0 | 3 | 0 | 2 | 2 | **13** |
| AIEd Bloom (2024) | 3 | 2 | 0 | 3 | 0 | 2 | 1 | **11** |

*Note: Almatrafi & Johri (2025) is excluded as it applies existing Bloom categories via automation rather than proposing a taxonomy. *

## 8.3 Interpretation

Three findings warrant discussion. ACF uniquely achieves maximum scores on both Completeness and Appropriateness-to-Purpose. Under our meta-characteristic, Completeness requires coverage of individual cognition, distributed cognition, and mode-governance as distinct assessment targets. No other framework systematically addresses all three. Appropriateness-to-Purpose requires generating assessable learning outcomes with mode-specific verbs. ACF's dual-mode structure with distinct verb pairs (Retrieve/Curate, Explain/Discriminate, Execute/Specify & Verify, etc.) and its tri-rubric assessment architecture directly serve this purpose.

Original Bloom achieves the highest total score due to Evidence of Usefulness. Seven decades of validated application across disciplines yields EV = 4. ACF scores EV = 1 because it is newly proposed. This limitation is acknowledged; however, EV rewards historical use rather than fitness for emerging contexts. A framework optimised for individual cognition cannot score highly on Completeness when the meta-characteristic requires coverage of distributed cognition.

Two frameworks score zero on both Completeness and Appropriateness-to-Purpose. AIEd Bloom (Hmoud & Shaqour, 2024) and Hammad et al. (2024) classify AI technologies rather than human cognitive processes. While valuable for instructional design and technology review respectively, they cannot generate human learning outcomes under our meta-characteristic. This underscores the importance of explicit meta-characteristic specification; a framework may be well-constructed for one purpose while unsuited for another.

The evaluation confirms the structural pattern identified in Section 4.2. Frameworks mapping AI capabilities against Bloom levels (Oregon State, Nehru et al.) preserve Lucidity and Parsimony but score moderately on Completeness because they do not distinguish distributed cognition as a separate mode. Frameworks adding AI-specific competencies (Gonsalves, AlAfnan) improve Completeness but introduce Orthogonality challenges. Frameworks abandoning hierarchy (Jain & Samuel) sacrifice the sequencing logic reflected in reduced Parsimony.

ACF's approach (retaining Bloom's levels while transforming each into dual-mode operation with Orchestration as distinct meta-cognitive capacity) achieves high Completeness and Appropriateness-to-Purpose while maintaining acceptable Lucidity and Orthogonality. The trade-off is reduced Parsimony (13 categories versus Bloom's 6) and untested Evidence of Usefulness.

These trade-offs are appropriate given the framework's purpose: addressing an assessment utility gap that more parsimonious frameworks cannot fill.

Oregon State University's *Bloom's Taxonomy Revisited* scores highly under the assessment-utility criteria used in this review, particularly in its practical guidance for assessment redesign and its articulation of human-centered skills alongside AI-capable tasks. This result should be taken at face value: OSU provides a strong assessment-facing response to generative AI within a largely familiar Bloom architecture. However, a high score on assessment-utility does not imply that the framework represents the multidimensional structure of cognition that motivates the present paper. OSU primarily treats AI as an external capability whose effects must be mapped onto Bloom levels. This is useful for identifying assessment vulnerabilities and redesigning tasks, but it does not explicitly model cognition as operating in two distinct modes (individual vs distributed) with mode-specific competencies and a governance layer for mode-switching, trust calibration, and degradation detection. On the present argument, this is not a defect so much as a difference in objective: OSU is best read as an assessment adaptation, whereas ACF is a taxonomy of dual-mode cognition and its regulation.

## 9. Conclusion

Cognition is now effectively enacted in two modes (individual and distributed) and educational taxonomies must represent this shift at the level of assessable learning outcomes. Most AI-era revisions retain individual cognition as the implicit norm by mapping AI capabilities onto Bloom, appending AI-related skills, or classifying platform functions. The Augmented Cognition Framework instead treats dual-mode performance as the organising principle of the taxonomy.

ACF specifies mode-distinct verbs for each traditional Bloom level, enabling instruction and assessment of both individual cognition and human–AI distributed cognition. It further models an asymmetric dependency in which distributed performance typically rests on individual foundations, while still accommodating structured scaffolding routes that temporarily invert sequencing under defined conditions. A seventh level, Orchestration, captures the governable competencies required to manage a stochastic, partially opaque external agent (mode selection, trust calibration, verification, and accountability) rather than internal self-monitoring alone.

The central contribution is assessment utility. The tri-rubric structure (Individual, Distributed, Orchestration) supports outcome specification and evaluation across three targets that existing frameworks do not separate systematically, and it remains implementable across institutional settings through flexible calibration.

These contributions also define a focused research agenda. First, ACF is currently conceptual and requires empirical validation, including discipline-specific calibration of thresholds and testing of the proposed sequencing and scaffolding claims. Second, ACF is limited to the cognitive domain; AI's implications for affective development and psychomotor competence warrant parallel taxonomic extensions. Third, implementation will be context- and time-sensitive: optimal mode balance and threshold settings will vary by discipline, role, and learner stage, and should be revisited as AI capabilities evolve. Together, these gaps do not weaken the framework's rationale; they specify the empirical and developmental work required to make dual-mode assessment robust at scale.